\documentclass[twocolumn,english,prl,showpacs,amsmath,amssymb,superscriptaddress,floatfix]{revtex4}

\makeatletter

\usepackage{times}
\usepackage{graphicx}

\usepackage{babel}
\makeatother
\begin{document}

\title{Spin qubits in double quantum dots - entanglement versus the Kondo effect}

\author{A. Ram\v{s}ak}

\affiliation{Faculty of Mathematics and Physics, University of Ljubljana, Ljubljana,
Slovenia}

\affiliation{J. Stefan Institute, Ljubljana, Slovenia}

\author{J. Mravlje}

\affiliation{J. Stefan Institute, Ljubljana, Slovenia}

\author{R. \v{Z}itko}

\affiliation{J. Stefan Institute, Ljubljana, Slovenia}

\author{J. Bon\v{c}a}

\affiliation{Faculty of Mathematics and Physics, University of Ljubljana, Ljubljana,
Slovenia}

\affiliation{J. Stefan Institute, Ljubljana, Slovenia}

\date{26 July 2006}

\begin{abstract}
We investigate the competition between pair entanglement of two spin
qubits in double quantum dots attached to leads with various
topologies and the separate entanglement of each spin with nearby
electrodes. Universal behavior of entanglement is demonstrated in
dependence on the mutual interactions between the spin qubits, the
coupling to their environment, temperature and magnetic field. As a
consequence of quantum phase transition an abrupt switch between fully
entangled and unentangled states takes place when the dots are coupled
in parallel.
\end{abstract}


\maketitle

{\it Introduction.}--
After the recent discovery of quantum computing algorithms, their
practical potential led to the interest in quantum entanglement
spurred on by the fact that if a quantum computer were built, it would
be capable of tasks impracticable in classical
computing\cite{nielsen00}. Nanostructures consisting of coupled
quantum dots are candidates for required scalable solid state arrays
of electron spin qubits\cite{loss98,divincenzo95}. The interaction of
such qubits with the environment is in general a complicated many-body
process and its understanding is crucial for experimental solid state
realisation of qubits in single and double quantum dots
(DQD)\cite{coish06}. 
Recent experiments on semiconductor double quantum dot
devices have shown that electron occupation may be controlled down to
the single-electron level by surface gates\cite{elzerman2003}. Also
spin entangled states were detected\cite{chen2004}, DQDs were used to
implement two-electron spin entanglement\cite{hatano2005}, and
coherent manipulation and projective readout\cite{petta05} was
demonstrated.

The purpose of entangled qubit pairs is to convey quantum information
through a computing device\cite{nielsen00}. The entanglement of two
spin qubits may be uniquely defined through von Neuman entropy or,
equivalently, concurrence\cite{bennett96,wootters1998}. A pair of
qubits may be realised, e.g., as two separate regions, each occupied
by one electron in a state $|s\rangle_{\mathrm{A},\mathrm{B}}$ of
either spin, $s=\uparrow$ or $\downarrow$. For a system in a
pure state
$|\Psi_{\mathrm{A}\mathrm{B}}\rangle=
\sum_{ss'}\alpha_{ss'}|s\rangle_{\mathrm{A}}\otimes|s'\rangle_{\mathrm{B}}$, 
the concurrence as a quantitative measure for
(spin) entanglement is given by\cite{wootters1998}
$C_0 = 2|\alpha_{\uparrow\!\downarrow}\alpha_{\downarrow\!\uparrow}-
\alpha_{\uparrow\!\uparrow}\alpha_{\downarrow\!\downarrow}|.$
Two qubits are completely entangled, $C_0=1$,
if they are in one of the Bell states\cite{bennett96}, e.g., singlet
$|\Psi_{\mathrm{A}\mathrm{B}}\rangle \propto {|\uparrow
\downarrow\rangle -|\downarrow \uparrow \rangle}$.

\begin{figure}
\begin{center}\includegraphics[width=50mm,keepaspectratio]{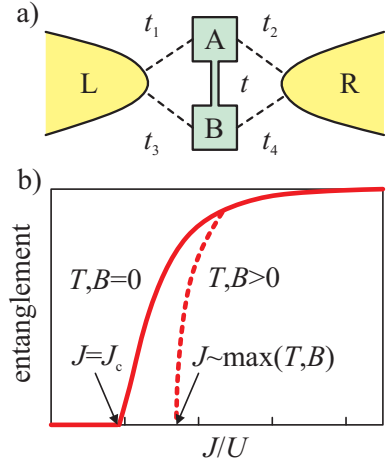}\end{center}
 
\caption{(color online)
(a) Double quantum dot system. $t$ and $t_{n}$ are the matrix elements
for tunneling between dots $\mathrm{A}$ and $\mathrm{B}$ and from the dots to the electrodes,
respectively.  (b) Entanglement between two spins
on the quantum dots as a function of the interdot exchange coupling
$J$: below $J_{c}$, the two spins are not
entangled and the DQD is in some type of the Kondo regime. For elevated temperatures
and magnetic field above $J_c$, the entanglement is zero if $J\lesssim \textrm{max}(T,B)$.}

\label{Fig1} 
\end{figure}

{\it The setup and main results.}--
We focus on entanglement between two electrons confined in two
adjacent quantum dots weakly coupled by electron tunneling in a
controllable manner, Fig.~\ref{Fig1}(a). The inter-dot tunneling matrix
element $t$ determines not only the tunneling rate, but also the
effective magnetic superexchange interaction $J\sim 4t^{2}/U$, where
$U$ is the scale of Coulomb interaction between two electrons confined
on the same dot. By adjusting a global back-gate voltage, exactly two
electrons can be confined to the dots $\mathrm{A}$-$\mathrm{B}$ on
average. 

Additional gate voltages are applied to independently control
tunneling to the electrodes, $t_{n}$. Depending on the values of
$t_{n}$, various topologies can be realised and even for very weak
coupling, the spin of confined electrons may be screened due to the
Kondo effect, where at temperatures below the Kondo temperature $T_K$ a
spin-singlet state is formed between a confined electron and
conduction electrons close to the Fermi energy.
Conductance and some other properties of such systems have
already been studied, without considering, however, the 
analysis of the entanglement and its relationship to the many-body
phenomena embodied in the Kondo effect.

Qualitatively the physics related to qubit pairs in coupled DQDs is 
plausible and can be summarised as:

(i) If $t/U$ is not small the electrons tunnel between the dots and
charge fluctuations introduce additional states with zero or double
occupancy of individual dots\cite{schliemann01, zanardi02}.  Due to
significant local charge fluctuations this regime is not particularly
appropriate for the spin-qubits manipulation.

(ii) For systems with strong electron-electron repulsion, charge
fluctuations are suppressed and the states with single occupancy --
the spin-qubits -- dominate. Due to the effective antiferromagnetic
Heisenberg interaction the spins $\mathrm{A}$ and $\mathrm{B}$ tend to
form the singlet state.  The physics of such qubit pairs may be
compared to the two-impurity Kondo problem studied by Jones, Varma and
Wilkins two decades ago\cite{jones1988}. There two impurities form
either two Kondo singlets with delocalised electrons or bind into a
local spin-singlet state which is virtually decoupled from delocalised
electrons. The crossover between the two regimes is determined by the
relative values of the exchange energy $J$ and twice the Kondo
condensation energy, of order the Kondo temperature $T_{K}$.

As shown in this paper, our numerical results for representative DQD
systems with different topology of possible experimental realisations
reveal much more diverse physical behaviour. However, (i) in all cases
the spin qubits are unentangled for $J$ below some critical value
$J_{c}$, where the actual value of $J_{c}$ crucially depends on the
setup topology, and (ii) at elevated temperatures $T>0$ and external
magnetic field $B\ne 0$ the entanglement is additionally suppressed
and generically zero when $J\lesssim\textrm{max}(J_{c},T,B)$, as
schematically shown in Fig.~\ref{Fig1}(b).

{\it Quantitative results.}--
For simplicity we model DQD using the two-site Hubbard Hamiltonian $H
= -t\sum_{s}(c_{\mathrm{A}s}^{\dagger}c_{\mathrm{B}s} +
c_{\mathrm{B}s}^{\dagger} c_{\mathrm{A}s}) +
U\sum_{i=\mathrm{A},\mathrm{B}}n_{i\uparrow}n_{i\downarrow}$, where
$c^\dagger_{is}$ creates an electron with spin $s$ in the dot
$i=\mathrm{A}$ or $i=\mathrm{B}$ and $n_{is}=c^\dagger_{is}c_{is}$ is
the number operator. The dots are coupled to the left and right
noninteracting lead as shown in Fig.~\ref{Fig1}(a).

DQDs as considered here can not be described with a pure quantum state and
concurrence is not directly given by $C_0$. It is related to the 
reduced density matrix of the DQD subsystem\cite{wootters1998,osterloh02,syljusen03}, 
where for systems that are axially symmetric in 
spin space the concurrence may conveniently be given in
the closed form\cite{ramsak06},
\begin{eqnarray}
C & = & \textrm{max}(0,C_{\uparrow\!\downarrow},C_{\parallel})
/(P_{\uparrow\downarrow}+P_{\parallel}),\label{eq:cmax}\\
C_{\uparrow\!\downarrow} & = & 2|\langle S_{\mathrm{A}}^{+}
S_{\mathrm{B}}^{-}\rangle|-2\sqrt{\langle P_{\mathrm{A}}^
{\uparrow}P_{\mathrm{B}}^{\uparrow}\rangle\langle P_{\mathrm{A}}^
{\downarrow}P_{\mathrm{B}}^{\downarrow}\rangle},\nonumber \\
C_{\parallel} & = & 2|\langle S_{\mathrm{A}}^{+}S_{\mathrm{B}}^{+}
\rangle|-2\sqrt{\langle P_{\mathrm{A}}^{\uparrow}P_{\mathrm{B}}^
{\downarrow}\rangle\langle P_{\mathrm{A}}^{\downarrow}
P_{\mathrm{B}}^{\uparrow}\rangle},\nonumber \end{eqnarray}
where $S_i^{+} = (S_i^{-})^{\dagger} = c_{i\uparrow}^{\dagger}
c_{i\downarrow}$ is the electron spin raising operator for dot
$i=\mathrm{A}$ or $\mathrm{B}$ and $P_i^{s}=n_{is}(1-n_{i,-s})$ is the
projection operator onto the subspace where dot $i$ is occupied by one
electron with spin $s$. $P_{\uparrow\downarrow} =\langle 
P^\uparrow_\mathrm{A} P^\downarrow_\mathrm{B} +
P^\downarrow_\mathrm{A} P^\uparrow_\mathrm{B}\rangle $ 
and $P_{\parallel} = 
\langle P^\uparrow_\mathrm{A} P^\uparrow_\mathrm{B} + P^\downarrow_\mathrm{A}
P^\downarrow_\mathrm{B}\rangle$ are probabilities for antiparallel and parallel
spin alignment, respectively.
 
We have determined concurrence for all three possible topologically
non-equivalent two terminal experimental arrangements: double quantum
dots (i) coupled in series, (ii) laterally side coupled, and (iii)
coupled in parallel. Concurrence was determined numerically from
Eq.~(\ref{eq:cmax}), where expectation values correspond to a
many-body state with the chemical potential in the middle of the
electron band, which guarantees that the dots are singly occupied,
$\langle n_{\mathrm{A},\mathrm{B}}\rangle =1$. Our extensive
investigation over the full parameter range for various topologies
indicates that all show generic behaviour outlined in
Fig.~\ref{Fig1}(b), but quantitatively can differ by many orders of
magnitude, which should be taken into consideration in experiments
with such DQD.  Numerical methods were based on the Gunnarsson-Sch\"
onhammer (GS)
projection-operator\cite{gunnarsson1985,mravlje2006,rr03} and
numerical renormalisation group (NRG)\cite{krishna1980a,
zitko06a, zitko06b} methods.

{\it Serially coupled DQD.}--
First we consider serially coupled DQD, which models entangled pairs
that may be extracted using a single-electron
turnstile\cite{hu05}. Here $t_{1,4}=t'$ and $t_{2,3}=0$ with the
hybridisation width of each dot $\Gamma=(t')^2/t_0$, where $4t_0$ is
the bandwidth of noninteracting leads. Entanglement of a qubit pair
represented with quantum dots in the contact with the leads (fermionic
bath) was not quantitatively determined so far, although this system
has already been extensively studied
\cite{georges1999,izumida2000,mravlje2006} (and references therein).

\begin{figure*}[htbp!]
\begin{center}
\includegraphics[  width=17cm,  keepaspectratio]{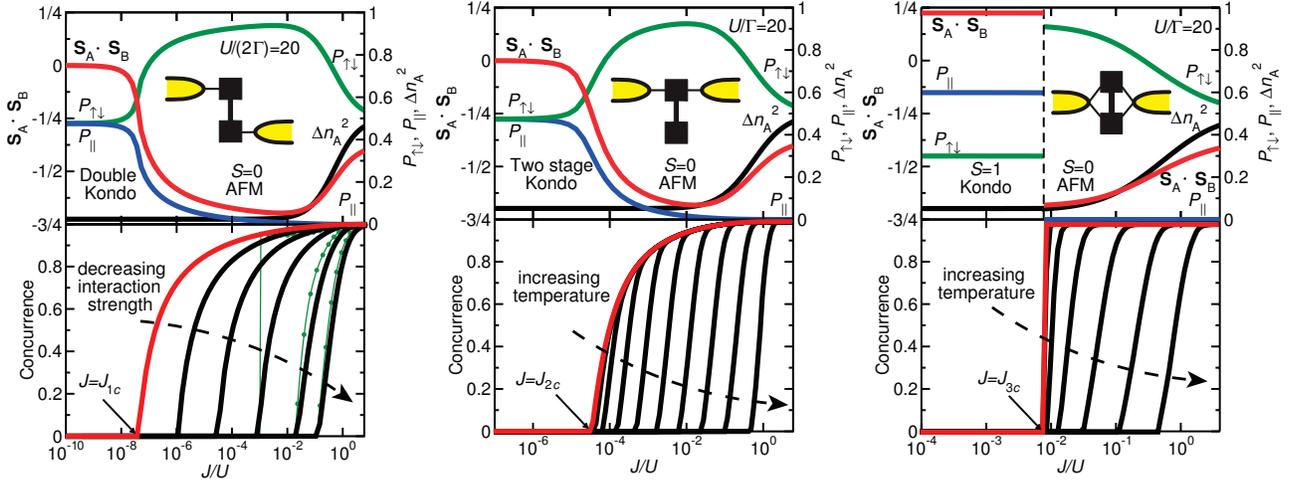}
\caption{(color online)
{\it Top panels:} Spin-spin
correlation function ${\it\bf S}_1 \cdot {\it\bf S}_2$, 
charge fluctuations on one quantum dot, $\Delta n_\mathrm{A}^2$, and
probabilities $P_{\uparrow\downarrow}$,  $P_{\parallel}$
for $T \ll T_K$, $B\to0$ 
and with $t^\prime=t_0/\sqrt{20}$. 
{\it Bottom left panel:} concurrence $C$, corresponding to serially coupled dots, for a range of  
interactions $U/\Gamma=40,32,24,16,8,4$ and calculated with both, the NRG and the GS
method (bullets), yielding the same $J_{1c}$, but
due to the limited span of variational basis the GS method progressively
overestimates $C$ for $U/\Gamma \gtrsim 20$ regime. 
{\it Bottom right two panels:} The results for side coupled and parallel
configuration obtained from the NRG method.
Temperatures range from the scale of the Coulomb repulsion parameter
$U$, $T/U=0.4$, to temperatures below the Kondo scale $T_K$; each
consecutive curve corresponds to a temperature lowered by a factor
4.
}
\label{Fig2}
\end{center}
\end{figure*}

In analogy with entanglement at zero temperature studied recently in a
many-body ground state\cite{jordan04}, we consider here concurrence of
DQD at fixed temperature and static magnetic field along the $z$-axis.  
Expectation values
$\langle ...\rangle$ in
the concurrence formula Eq.~(\ref{eq:cmax})
correspond to thermal equilibrium of the system, therefore $\langle
S_{\mathrm{A}}^{+}S_{\mathrm{B}}^{+}\rangle=0$ here. Qualitatively, the concurrence
is significant when enhanced spin-spin correlations indicate inter-dot
singlet formation. As shown in Fig.~\ref{Fig2}, the correlator
$\langle {\bf S}_\mathrm{A} \cdot {\bf S}_\mathrm{B} \rangle$ 
tends to $-3/4$ for $J$ large enough to suppress
the formation of Kondo singlets, but still  $J/U\ll1$, that
local charge fluctuations $\Delta n_\mathrm{A}^2$ are sufficiently suppressed 
and $P_{\uparrow\downarrow}+P_{\parallel}\to 1$.  Concurrence,
calculated for various values of the Coulomb interaction strengths and in 
the absence of magnetic field is presented in
Fig.~\ref{Fig2}, left bottom panel. 
As discussed above, $C$ is zero for $J<J_{1c}$ due to the Kondo effect,
which leads to entanglement between localised and conducting electrons\cite{adam06}
instead of the $\mathrm{A}$-$\mathrm{B}$ qubit pair entanglement. In finite magnetic field
irrespectively of temperature the 
concurrence abruptly tends to zero for $B>J$ (not shown here)\cite{magnetic}.

\begin{figure}

\begin{center}\includegraphics[  width=60mm,  keepaspectratio]{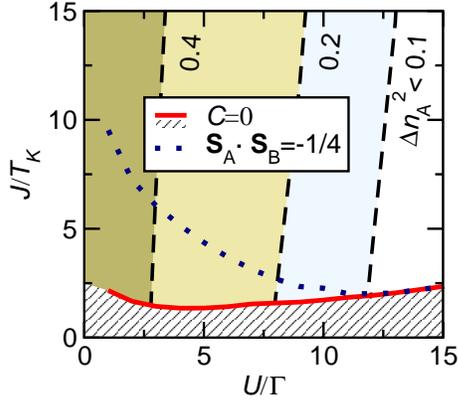}\end{center}
\caption{(color online)
In the phase diagram $(U/\Gamma,J/T_K)$ 
full line separates $C>0$ and $C=0$ regions
(line shaded), together with dotted line indicating $ \langle
\textrm{\bf S}_{\mathrm{A}}\cdot\textrm{\bf
S}_{\mathrm{B}}\rangle=-1/4$ ($T\to 0$ and $B=0$). Both lines merge for $U/\Gamma \gtrsim 12$, 
where charge fluctuations 
$\Delta n_\mathrm{A}^2$ progressively become negligible (contour plot).
\label{Fig3}}
  
\end{figure}

In particular, the local dot-dot singlet is formed and $C\ge 0$ whenever
singlet-triplet splitting $J>J_{1c}\sim2.5T_{K}(\Gamma)$, where the
Kondo temperature is given by the Haldane formula $T_K(\Gamma) =
\sqrt{U\Gamma/2} \exp(-\pi U/8\Gamma)$. This is presented in the phase
diagram in the $(U/\Gamma, J/T_K)$ plane, Fig.~\ref{Fig3}.  Dashed
region corresponds to the regime of zero concurrence and is delimited
by the line of the critical $J=J_{1c}$ (red line). The charge
fluctuations (Fig.~\ref{Fig3}, contour plot) are suppressed for
sufficiently large repulsion, i.e., $U/\Gamma \gtrsim 10$. In this limit and
in vanishing magnetic field, the DQD can be described in terms of the
Werner states\cite{werner89} and becomes similar to recently studied
problem of entanglement of two Kondo spin impurities embedded in a
conduction band\cite{cho06}. In this case,
$C_{\uparrow \downarrow}\sim 2(-\langle\textrm{\bf
S}_{\mathrm{A}}\cdot\textrm{\bf
S}_{\mathrm{B}}\rangle-\frac{1}{4})\sim P_{\uparrow
\downarrow}-2P_{||}$ for $C_{\uparrow \downarrow}\geq 0$. For large
$U/\Gamma$, where the charge fluctuations vanish, the
$\langle\textrm{\bf S}_{\mathrm{A}}\cdot\textrm{\bf
S}_{\mathrm{B}}\rangle=-\frac{1}{4}$ boundary (Fig.~\ref{Fig3}, dotted
line) progressively merges with the $C=0$ line.

{\it Side-coupled DQD.}--
In the side-coupled DQD configuration, $t_{1,2}=t'$, $t_{3,4}=0$
(Fig.~\ref{Fig2}, middle), the dot $\mathrm{A}$ is in direct contact with
the electrodes, while the dot $\mathrm{B}$ couples to the conduction band
only indirectly through the dot $\mathrm{A}$. Because the two electrodes are 
in contact only with dot $\mathrm{A}$, is
$\Gamma=2(t')^2/t_0$, i.e. twice as much as in the previous case. Since $T_K
\propto \exp(-\pi U/8\Gamma)$, the Kondo temperature on dot $\mathrm{A}$ is
strongly enhanced.

For $J>J_{2c}$, the spins bind in an antiferromagnetic singlet, as in all
other cases. For $J<J_{2c}$, the system enters the 'two stage Kondo' regime,
characterised by consecutive screening of local moments
\cite{cornaglia2005tsk, zitko06a, vojta2002}. At the Kondo temperature
$T_K^{(1)}=T_K(\Gamma)$ the spin on the dot $\mathrm{A}$ is screened, while
the spin on the dot $\mathrm{B}$ is compensated at a reduced temperature
\begin{equation}
T_K^{(2)}=d_1 T_K^{(1)} \exp(d_2 J/T_K^{(1)}),
\label{tk2}
\end{equation}
where $d_{1,2}$ are constants of order unity. The Kondo effect on dot $\mathrm{A}$
leads to the formation of a local Fermi liquid for temperatures below
$T_K^{(1)}$. The quasiparticle excitations of this Fermi liquid then
participate in the Kondo effect on dot $\mathrm{B}$ at much lower Kondo temperature
$T_K^{(2)}$\cite{cornaglia2005tsk}. Such description is valid only
when the temperature scales $T_K^{(1)}$ and $T_K^{(2)}$ are widely
separated. This no longer holds when $J$ becomes comparable to $T_K^{(1)}$,
see Eq.~\eqref{tk2}. The critical $J_{2c}$ is thus still given by $J_{2c}
\sim T_K^{(1)}$. The crossover is very smooth and the transition from the
inter-impurity singlet phase to the Kondo phase does not exhibit any sharp
features. In fact, the low temperature fixed point is the same for
$J<J_{2c}$ and $J>J_{2c}$, unlike in the case of DQD in series. In the
latter case, the Kondo phase and the inter-impurity singlet phase are
qualitatively different and are characterised by different electron
scattering phase shifts.

In the $J>J_{2c}$ singlet region, when the temperature is above $J$, the
exchange interaction is too weak to bind the spins into a singlet and the
entanglement is lost (see Fig.~\ref{Fig2}, bottom middle panel). In the $J<J_{2c}$
Kondo region, the concurrence is zero irrespective of temperature: for
$T<T_K$ it is zero due to the Kondo effect, and for $T>T_K$ the spin-singlet
cannot be restored, since $T>J$. Elevated temperature and magnetic field 
dependence is similar to the previous
case of serially coupled dots\cite{magnetic}. 

{\it Parallelly coupled DQD.}--
In the case of parallel quantum dots ($t_{n} \equiv t'$ and $\Gamma=2(t')^2/t_0$)
the physics is 
markedly different from the case of the previous two configurations. 
The conduction band mediated effective
Ruderman-Kittel-Kasuya-Yoshida (RKKY) interaction 
between the dots is, in our simplified model, ferromagnetic
\cite{zitko06b}. Here the spins order ferromagnetically
into a triplet state (Fig.~\ref{Fig2}, right panels) and undergo a $S=1$
Kondo screening at low temperatures in the regime  
\begin{equation}
J < |J_\mathrm{RKKY}| \sim c \frac{64}{\pi^2}
\frac{\Gamma^2}{U}, 
\end{equation}
where $c$ is a constant of order unity. This yields an uncompensated
$S=1/2$ residual spin and since half a unit of spin is quenched by the
conduction band via the Kondo effect, there clearly cannot be any
entanglement between the electrons on the dots. For $J >
|J_\mathrm{RKKY}|$, the antiferromagnetic ordering wins and, once
again, the electrons form an entangled singlet state. The transition
from the Kondo phase to the singlet phase is in this case a true
quantum phase transition\cite{zitko06b}, and the concurrence drops
abruptly from (surprisingly) $C\approx 1$, to $C=0$ when the exchange
$J$ between the dots is decreased below some $J_{3c}$.  Due to the
sensitivity of concurrence on $J\sim J_{3c}$ can therefore
parallely coupled DQD
be considered a perfect entanglement switch.  Critical coupling
$J_{3c}$ is not determined by $T_K$, as previously, but rather by
$|J_\mathrm{RKKY}|$. For this reason, concurrence drops to zero at a
much higher temperature (compare middle and right bottom panels in
Fig.~\ref{Fig2}), and is strictly zero for $J\lesssim T$ as in
previous two cases.  In finite magnetic field $C=0$ if $J\lesssim
|J_\mathrm{RKKY}|+B$ (not shown here)\cite{magnetic}.

If the couplings $t_n$ are not strictly equal, another Kondo screening stage
may occur at low temperatures, in which the residual $S=1/2$ spin is finally
screened to zero\cite{jayaprakash1981}. In this case the quantum phase
transition is replaced by a cross-over that becomes smoother as the degree
of asymmetry between the couplings $t_n$ increases (results not shown).

{\it Conclussions.}--
We have found generic behaviour of spin-entanglement of an electron
pair in double quantum dots. On the one hand, we have shown
quantitatively that making the spin-spin exchange coupling $J$ large
by increasing tunneling $t$, leads to enhanced charge fluctuations,
whilst on the other, a small interaction $J<J_{c}$ suppresses
entanglement as the DQD system undergoes some form of the Kondo
effect.  Various regimes are explained analytically and supported with
typical numerical examples. In the limiting cases we found (i) two
separate Kondo effects for serially coupled DQD; (ii) two-stage Kondo
effect in side-coupled DQD; and (iii) $S=1$ Kondo effect with
underscreening for parallelly coupled DQD, eventually followed by
another $S=1/2$ Kondo effect at lower temperatures. For two terminal
setups, these are the only possible types of the Kondo effect; in a
generic situation with arbitrary $t_n$, one of these possibilities
must occur.

In all cases, in spite of different Kondo mechanisms, the  temperature
and magnetic field dependence of entanglement is proven to be
determined solely by the exchange scale $J$ and not by the much lower
scale of the Kondo temperature, which explains the universal behaviour
of the entanglement shown in Fig.~\ref{Fig1}(b). Critical $J_c$, however,
will for various experimental setups vary for several orders of magnitude.

We thank I. Sega for inspiring discussions and J.H.  Jefferson for
valuable comments regarding the manuscript. One of the authors (A.R.) 
acknowledges a helpful discussion with 
R.H. McKenzie.  We acknowledge support from the Slovenian Research
Agency under contract Pl-0044.


\end{document}